\documentclass[nohyper]{JHEP}
\usepackage{amsmath}
\usepackage{amssymb}
\usepackage{cite}
\usepackage{xspace}

\usepackage{epsfig}

\psfull

\newcommand{\ie}{\emph{i.e.}\ }
\newcommand{\eg}{\emph{e.g.}\ }

\def\wrt{with respect to }

\def\Em{E_{\min}}

\def\be{\beta}

\def\nf{n_{\!f}}

\def\as{\alpha_{{\textsc{s}}}}

\def\Et1{E_{t1}}
\def\Et2{E_{t2}}

\def\Emin{E_{\min}}

\title{Dijet rates with symmetric $E_t$ cuts.}%
\author{Andrea Banfi\\
NIKHEF Theory group, P.O. Box 41882, 1009 DB Amsterdam, The
Netherlands.\\
\email{andrea.banfi@nikhef.nl}
}
\author{Mrinal Dasgupta \\
  CERN, Theory Division, CH-1211 Geneva 23, Switzerland.\\
\email{mrinal.dasgupta@cern.ch}}

\preprint{
hep-ph/0312108 \\
CERN-TH/2003-227\\
NIKHEF-2003-016
}  

\abstract{We consider dijet production in the region where symmetric
  cuts on the transverse energy, $E_t$, are applied to the jets. In
  this region next-to--leading order calculations are unreliable and an all-order resummation of soft gluon effects is needed, which we carry out. Although, for illustrative purposes, we choose dijets produced in deep inelastic scattering, our general ideas apply additionally to dijets produced in photoproduction or $\gamma \gamma$ processes and 
should be relevant also to the study of prompt di-photon $E_t$ spectra 
in association with a recoiling jet, in hadron-hadron processes.}


\begin{document}

\section{Introduction}

Measurements involving jet production and comparisons of the
corresponding rates and distributions with QCD calculations have
provided some of the best means for testing perturbative QCD.  As an
example, final states involving two or more jets have been extensively
studied at HERA by the H1 and ZEUS collaborations~\cite{H1,ZEUS} over
a wide kinematic range.  At high photon virtualities $Q^2$,
comparisons of dijet cross-sections and distributions with
next-to--leading order (NLO) calculations~\cite{FrixRid,Klasen,DISENT,DISASTER,NLOJET,Jetvip} have yielded 
precise measurements
of the strong coupling $\alpha_s$~\cite{HERA}.  On the other hand
low $Q^2$ dijet measurements have been used as means of obtaining
information on the gluon distribution $xg(x)$ of the proton,
complementary to that obtained from structure function scaling
violations~\cite{ZEUS}.  However such studies are by no means unique to the HERA
experiment and dijet production has been actively studied for $\gamma
\gamma$ collisions at LEP~\cite{LEP} and $pp$ collisions at the
Tevatron~\cite{Tevatron}.

A feature common to most experimental analyses on dijets is the
presence of selection cuts which define the phase space for jet
production and are generally meant to ensure that the kinematic region
chosen is least affected by theoretical uncertainties.  In
experimental analysis of inclusive dijet final states, one usually
imposes cuts on the transverse energy $E_t$ of each of the two highest
$E_t$ dijets.  It was observed some time ago by Frixione and Ridolfi
\cite{FrixRid} that NLO calculations for dijet rates break down if
symmetric $E_t$ cuts $E_{t1},E_{t2} \geq \Emin$ are used on
the two highest $E_t$ jets.  The breakdown of the NLO calculation was
observed to occur because of sensitivity to soft gluon emission, which
subsequently led to the region of symmetric cuts being dubbed infrared
unsafe or unphysical.

Frixione and Ridolfi suggested the asymmetric cuts $E_{t2} \geq
E_\mathrm{{min}}$, $E_{t1} \geq E_\mathrm{{min}}+\Delta$, with
$\Delta$ not too small compared to the hard scale of the process,
which reduced the sensitivity to soft gluon effects and resulted in
more reliable NLO predictions. Experimental data, on the other hand, can
accurately be obtained even in the presence of symmetric cuts.  For
example the cross-section for dijet production in deeply inelastic
scattering (DIS) has been experimentally measured over a range of
values of $\Delta$, down to the symmetric region $\Delta =0$
(see \eg Ref.~\cite{ZEUS}).

A plot of the data for $\sigma(\Delta)$, the dijet cross-section with
a given value of $\Delta$ (keeping $E_{\mathrm{min}}$ fixed), versus
$\Delta$ can be found in~\cite{ZEUS} and illustrates the problem
clearly.  It shows that the total rate $\sigma(\Delta)$ is a
monotonically decreasing function of $\Delta$ with its maximum value
$\sigma(0)$ corresponding to the rate with symmetric cuts,
$\Delta=0$.  This is clearly expected on the basis of simple phase
space considerations: increasing $\Delta$ means that less of the total
phase space is available and the dijet rate falls.

The NLO calculation (the program DISENT~\cite{DISENT} was used for
this purpose in Ref.~\cite{ZEUS}) on the other hand, agrees with
the data for larger $\Delta$ values, but as one lowers $\Delta$ there
is a turnover of the NLO calculation and the corresponding curve
starts to fall, whilst the data rises continuously.  At $\Delta=0$ in
particular there is a significant difference between the data and the
NLO estimate $\sigma_{\mathrm{NLO}}(0)$.  Hence at the very point
where the measured cross section is largest there is a maximal
discrepancy of the NLO result with the data, and in the vicinity of
this point, a qualitative behaviour different from that indicated by
the data.  Therefore quite clearly, a better understanding is sought of
the theoretical limitations that lead to a breakdown of the NLO
estimate at small $\Delta$.

Moreover the problem discussed above is quite general. It also appears
when one considers, for example, the hadroproduction of a prompt
photon in association with a jet. The corresponding fragmentation
contribution (when the jet emits a hard collinear photon) is important
as a background for Higgs searches.  Here once again, placing symmetric
cuts on the final state photon and jet $E_t$ values will lead to
infrared sensitivity of the NLO calculation.  Alternatively one can
consider prompt diphoton production in association with a
jet and study the photon pair $E_t$
distribution~\cite{DelDuca}. Putting a cut on the recoiling jet $E_t$ one can
investigate the $E_t$ distribution of the di-gamma pair.  Doing this
in NLO QCD, an unphysical discontinuity arises at the position of the
cut, due to fact that in that region soft gluon emission becomes
important. For this paper however we shall continue to
use dijet production in DIS as our illustrative example and will
consider extensions to other processes in forthcoming work.

In the present paper we point out that the unphysical behaviour in the
dijet rate near $\Delta = 0$ is due to the presence of large
logarithms of $Q/\Delta$ (where $-Q^2$ is the photon virtuality) in the
slope $\sigma'(\Delta)=d\sigma/d\Delta$. The logarithms in question
arise from a veto on real gluon emission above scale $\Delta$,
effective in a certain part of the dijet phase space, which causes
uncanceled virtual corrections to build up above this scale.

While one will obtain double logarithms (two powers of $\ln Q/\Delta$
for every power of $\alpha_s$) from emissions soft and collinear to
the incoming parton, one will obtain single logarithms
$\alpha_s^n\ln^n Q/\Delta$ from soft gluon emission at large angles.\footnote{In several common  jet algorithms, \eg  cone algorithms and
  their variants, and the inclusive $k_t$ algorithm~\cite{CDMW,ElSop}, there
  will be no soft  and collinear double logarithmic (DL) contributions from
  the outgoing hard partons, provided one recombines partons into jets
  appropriately,  which we  shall discuss  shortly.}  
These logarithms
cause the slope  calculated at NLO, $\sigma'_{\mathrm{{NLO}}}(\Delta)$, to change
sign becoming positive, at small $\Delta$, and divergent at  
$\Delta=0$. This  property of  the slope is    reflected as a  leading
$\Delta  \alpha_s \ln^2 \Delta $  term in the  NLO computation for the
total  rate   $\sigma(\Delta)$  at     small  $\Delta$.   Thus   while
$\sigma(\Delta)$ has a {\it{finite}} value at $\Delta = 0$,  this value is not
correctly given by any fixed order of  perturbation theory.  One needs
to first resum the large  logarithms in the slope $\sigma'(\Delta)$, to
all  orders,   to   obtain  a    physically  meaningful   result    for
$\sigma(\Delta)$, at small $\Delta$.

In this paper we resum soft gluon effects (including hard collinear
emission from the incoming leg) to all orders in perturbation theory to
account for the above large logarithms in the slope
$\sigma'(\Delta)$. Our resummation will be in the space of a
Fourier variable $b$ conjugate to $\Delta$ and we shall resum
logarithms in $b$ that, at large $b$, reflect the singular behaviour at
small $\Delta$.  These logarithms shall be resummed into a {\it{form
    factor}} $\Sigma$, which can be expressed as
\begin{equation}
\ln \Sigma(b) = Lg_1(\alpha_s L) +g_2(\alpha_s L),
\end{equation}
with $L \equiv \ln(bQ)$ and $g_1$ and $g_2$ being the leading and
next-to--leading logarithmic functions. We shall refer to this as
next-to--leading logarithmic (NLL) or single-logarithmic (SL) accuracy.

Another factor we have to consider however is the conservation of
transverse momentum. Vectorial cancellations between harder emissions
are another way of obtaining a small $E_t$ difference between the
final state jets and this effect also impacts the slope
$\sigma'(\Delta)$ at very small $\Delta$. The full answer will be
given by the convolution of the form factor $\Sigma$ with an oscillatory
function: 
\begin{equation}
  \label{eq:cancellation}
\sigma'(\Delta) \sim \int_0^{\infty}\frac{db}{b} \sin(b\Delta) \Sigma(b),   
\end{equation}
where the sine function accounts for vectorial transverse momentum
conservation.  Once this convolution is performed to obtain the
resummed slope one finds that the unphysical behaviour is replaced by
a smooth behaviour in the limit $\Delta \to 0$. In fact instead of
diverging the slope remains finite (and negative as is physically
required) and for $\Delta/Q \ll 1$ 
is proportional to
$\Delta$, \ie is linear.  

To obtain the maximal possible accuracy, 
one has to {\it{match}} the resummation performed here, 
with fixed order computations 
that account for subleading logarithms and finite corrections 
(constant pieces
and non-logarithmic terms in $\Delta$). These would start at NLO and will be
important to get a good description 
at larger $\Delta$ values. 
We shall postpone dealing with the
issue of matching to subsequent work.

In all of the above considerations, the definition of jets will
naturally have a significant impact on the answer. One has to choose
both a jet algorithm and a recombination scheme which details how the
kinematic properties of the jet (such as its $E_t$) relate to those
of its partonic constituents. The results we present here are based on
the use of a cone-type algorithm also used previously in theoretical
studies involving dijets~\cite{KidSter}. We shall mention some details
of this subsequently. 
We require also that particles are clustered into jets using a
four-vector recombination scheme $p_{\mathrm{jet}} = \sum_{i \in
  \mathrm{jet}} p_i$ where $p$ labels four-momenta and the sum runs
over all partons that constitute the jet.  With this recombination
scheme the leading logarithmic function $g_1$ is independent of the
details of the jet algorithm, since it arises purely from emissions collinear
to the incoming parton, which will not be recombined with the outgoing
jets \footnote{What we actually need, for our calculations to directly apply, is a recombination scheme that vectorially adds the three-momenta of partons within a jet. Then the jet $E_t$ is just defined as the magnitude of the corresponding jet transverse-momentum vector $\vec{p}_{t,{\mathrm{jet}}}$.} 
Hence the specifics of the jet algorithm enter the function $g_2$,
\ie at NLL level.

The outline of this paper is as follows. In the next section we shall consider the situation at leading (Born) order and introduce all the quantities involved.  In the following section we  illustrate the origin of the soft gluon problem at NLO in more detail and calculate the  DL divergence at NLO.  Next we shall perform the all-order resummation to NLL accuracy, in $b$ space.  Subsequent to this we shall present our final numerical results and a discussion illustrating the main features thereof. Lastly we shall conclude while mentioning some planned future developments and work in progress.

\section{Dijet production at leading order}
\label{sec:dijet}
Let us consider the production of two final state jets in the Breit frame of 
DIS.  To leading order, the dijets are just two
partons labeled $k$ and $r$ (see figure~\ref{fig:Born}). We also denote
with $p$ the incoming parton four-momentum and with $q$ the
four-momentum of the virtual photon.  Further one imposes the
asymmetric cuts
\begin{equation}
E_{t1} > E_{\min} + \Delta\>,\qquad E_{t2} > E_{\min}\>,
\end{equation}
where $E_{ti}$ denotes the transverse energy of the $i^{th}$ jet (and
to this order parton $r$ or $k$) with respect to the photon axis in
the Breit frame.

\FIGURE{
  \epsfig{file=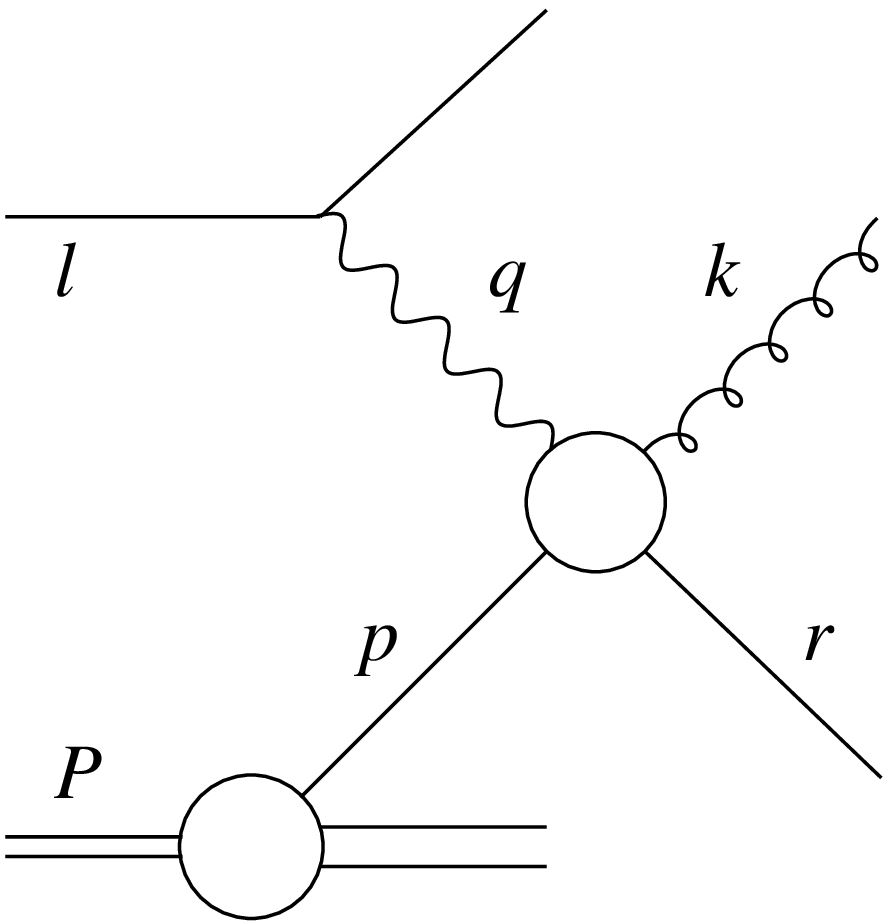, width=.6\textwidth}
  \caption{The Born kinematics of dijet production in DIS.}
  \label{fig:Born}}  

The general expression for the total rate (with the above cuts
imposed) for dijet production in $x$ space can be written at leading
order as
\begin{equation}
\sigma_{0}(x,Q,\Emin,\Delta) =  \int_x^{1} \frac{d\xi}{\xi}
C_{0} (\xi,Q,\Emin,\Delta) f \left (\frac{x}{\xi},Q^2\right )\>,
\end{equation} 
where the subscript '0' denotes the fact that the above result is at
Born level, $x$ denotes the usual DIS Bjorken variable while $\xi =
Q^2/(2(p\cdot q))$ is the fraction of the incoming parton momentum $p$
carried by the struck parton.  In the above formula we
have used $C_{0}$ to denote a generic coefficient function,
implicitly including in it the transverse and longitudinal components
(for simplicity we confine the discussion to virtual photon exchange
only). Accordingly $f(x/\xi)$ has been used to
denote the parton density and 
includes the dependence on parton flavour and charge. We shall always consider the renormalisation scale and the factorisation scale as equal to $Q$, but variations around this value can be trivially accounted for.

The leading order coefficient function $C_0$ can be obtained by
integrating the squared matrix element $M^2$ over the desired phase
space as below:
\begin{multline}
\label{eq:born}
C_0{(\xi,\Delta)} = \int \frac{d^3 \vec{r}}{2(2\pi)^3 r_0}
\frac{d^3 \vec{k}}{2(2 \pi)^3 k_0} (2 \pi)^4 \delta^4(p+q-k-r)
M^2(p,q,r,k) \\ \times \Theta(E_{t1} -(\Emin+\Delta)) \Theta(\Et2 -\Emin)
\>.
\end{multline}
Here $M^2(p,q,r,k)$ is the leading order matrix element squared for the hard dijet production at leading order and is made up of invariants constructed from the various parton momenta.  It differs according to whether the subprocess we consider involves an   incoming quark or gluon but the general form above applies in both cases.  We shall avoid displaying this dependence explicitly as well as the dependence on $\Emin$ and $Q$, in what follows below. In \eqref{eq:born} $r_0$ and $k_0$ denote the final-state particle energies.

Integrating away various components, using the delta function and
noting that at Born level $|\vec{r}_t| = r_t =|\vec{k}_t| =
E_{t1}=\Et2$, we are left with
\begin{equation}
C_0(\xi,\Delta) = \int \frac{d^2\vec{r}_t}{4 (2 \pi)^2}\frac{1}{|r_0
k_3 - k_0 r_3|} \Theta(r_t - (\Emin+\Delta))\; M^2,
\end{equation}
where $r_3$ and $k_3$ are the corresponding 
parton longitudinal momentum components along 
the photon axis, now fixed in terms of the components of $\vec{r}_t$, 
 and we identified each outgoing parton with a jet.

We now introduce the slope $\sigma'(\Delta) = d\sigma/d\Delta$. At leading order this is just
\begin{equation}
\label{eq:slopeint}
\sigma'_{0}(x,Q,\Emin,\Delta) = \int_x^{1} \frac{d\xi}{\xi}
C_{0}^{'}(\xi,Q,\Emin,\Delta) f\left (\frac{x}{\xi},Q^2\right )\>,
\end{equation} 
with the coefficient function obtained by differentiation of \eqref{eq:born} \wrt $\Delta$,

\begin{equation}
\label{eq:slope-Born}
C'_0(\xi,\Delta) = \frac{\partial C_0}{\partial \Delta} =
-\int \frac{d^2\vec{r}_t}{4 (2 \pi)^2}\frac{1}{|r_0 k_3 - k_0 r_3|}
\delta(r_t - (\Emin+\Delta))\; M^2.
\end{equation}

This integral can be performed (with any additional cuts such as one
on the interjet rapidity) and has a finite value as $\Delta \to 0$. Moreover at this order 
the slope is negative at all $\Delta$ as one requires on physical grounds. 
As we shall see this is no longer the case at NLO.


\section{Soft gluon effects at NLO}
The aim of this section will be to discuss the kinematical constraint on soft gluon emission, that arises in the region of small $\Delta/Q$, when one moves beyond the leading order eq.~\eqref{eq:slopeint}.
This constraint results in logarithmic enhancements and we shall explicitly compute the DL behaviour $\alpha_s \ln^2 Q/\Delta$, that first arises at NLO. Before that we discuss the relevant kinematics.
\subsection{Kinematics}
Moving to NLO we have to treat additional gluon emission. 
If the additional gluon is recombined with an outgoing hard parton by 
a jet algorithm and with four vector addition, it does not cause a mismatch in the jet transverse energies, $E_{t1}=E_{t2}$, and the jets are exactly 
back-to--back in azimuth. 
In this region the soft gluon contribution cancels with virtual corrections.
If however the gluon is not recombined into an outgoing jet, \eg it is near the beam (incoming parton) direction, it contributes to a transverse 
energy mismatch between the jets. 
Now the soft gluon effects do not cancel fully 
with virtual corrections and large logarithms appear.

To explicate this, we write the four-momenta of the outgoing partons as (here we explicitly identify partons $r$ and $k$ with jets and assume $k'$ is not recombined with them):
\begin{align}
\vec{r}_t & = E_{t1} (1,0) \>,\\
\vec{k}_t &= \Et2 (\cos(\pi \pm \epsilon), \sin(\pi \pm \epsilon))\>,\\
\vec{k'}_t &= k'_t (\cos\phi,\sin\phi)\>,
\end{align}
where the two jets are almost back-to-back in the transverse plane, since
if $k'$ is soft the hard parton recoil $\epsilon$ is small.  
From transverse
momentum conservation one gets
\begin{equation} 
k_t^{\prime 2} = E_{t1}^2+\Et2^2 -2 E_{t1} \Et2 + \epsilon^2 E_{t1} \Et2 +
\mathcal{O} (\epsilon^4 E_{t1} \Et2)\>.
\end{equation}
Additionally using $\epsilon^2 = \left ( k_t^{\prime 2}/\Et2^2 \right)
\sin^2 \phi$ (assuming $E_{t1} \approx \Et2$ in the soft emission
limit, discarding subleading terms and terms that are important over
only a parametrically small interval in $\phi$) one has simply that
\begin{equation}
\label{eq:mismatch}
k'_t |\cos\phi| = |k_x'|= |E_{t1} - \Et2|\>.
\end{equation}

The terms we neglected will not contribute at the NLL accuracy we aim for in this article. Thus at our level of accuracy, the mismatch in jet $E_{t}$ arises from a particular component of soft gluon transverse momenta (the component along the jet axis in the plane transverse to the photon axis).
Now that we have established how precisely soft gluons flowing outside the jets contribute to an $E_t$ mismatch between them, we can consider what happens due to the placement of $E_t$ cuts on the high $E_t$ dijets.

\FIGURE{
  \epsfig{file = 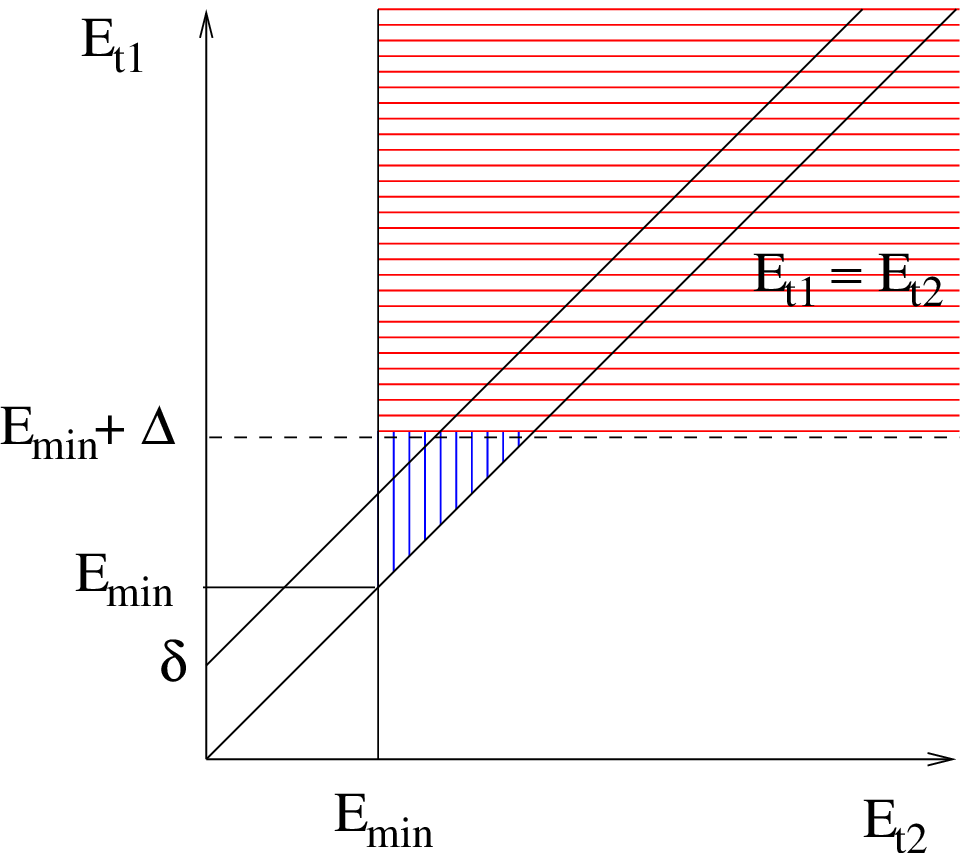}
  \label{fig:phasdiag}
  \caption{Phase space diagram in the $E_{t1},E_{t2}$ plane.}}

In this regard it is helpful to examine the diagram in
figure~\ref{fig:phasdiag}, which depicts the situation in the
$E_{t1}$, $E_{t2}$ plane. The shaded rectangular region shown, corresponds
to $E_{t1} \geq \Emin+\Delta$, $E_{t2} \geq \Emin$ and hence denotes the
region allowed by the experimental cuts. Let us consider the 
contribution from points along the dotted line $E_{t1}\!=\!\Emin+\Delta$, which contributes to the slope as is evident from \eqref{eq:slope-Born}. 
At Born level additionally, we are kinematically
constrained to be on the line $E_{t1} = E_{t2}$ (the lower of the
diagonal lines in the figure) and hence the contribution we obtain is
from the single point $E_{t1}\! =\! E_{t2} \!=\!\Emin+\Delta$. 

Now consider
the emission of a soft gluon $k'$ 
that causes a small mismatch $E_{t1} =
E_{t2}+\delta$, depicted by the upper of the diagonal straight lines in
the plot.  For this soft gluon to contribute, it must intersect the
dashed horizontal line $E_{t1} = E_{\mathrm{min}}+\Delta$, {\it{inside}} the allowed region (and hence
must pass through the shaded triangular region). If the soft
gluon has energy  (more precisely as we discussed the component $k_x$) 
$\delta\geq \Delta$ then it
is pushed outside the allowed region and hence vetoed. A veto on soft
emissions above some small scale causes uncanceled virtual
contributions at that scale, which in turn leads to logarithmic
behaviour in $\Delta/Q$. This behaviour will be manifest in the
derivative of the total rate \wrt $\Delta$, which receives
contributions only from points on the dotted line $E_{t1}
=\Emin+\Delta$. 

Such a DL behaviour in the slope of
the dijet rate (plotted against $\Delta$) is present in the fixed
order computations with DISENT~\cite{DISENT} and causes an
unphysical turnover of the theoretical calculation as we move to the
small $\Delta$ region. A resummation of soft gluon effects is thus
required to restore the correct physical behaviour seen for example in
the ZEUS data~\cite{ZEUS}.

\subsection{Soft one-loop calculation}
We shall now compute explicitly the above described soft gluon behaviour 
at NLO (one-loop).  
Adding a real soft gluon with four momentum $k'$ (with
energy $\omega' \ll Q$) to the Born system (which in both
incoming quark and gluon channels is a configuration with three hard
partons, two fermions and a gluon), we
have the real contribution to the 
NLO coefficient function\footnote{We have neglected the dependence of the parton
  distributions on the soft gluon $k'$ which we are allowed to do for soft emissions . We shall correct for this
  effect when including next-to--leading (single) logarithms arising
  from hard emissions collinear to the initial state partons, where  one cannot make this simplification.} 
\begin{multline}
\label{eq:nlo}
C_{1,{\mathrm{real}}}(\xi,\Delta) \approx \int \frac{d^3
  \vec{r}}{2(2\pi)^3 r_0} \frac{d^3 \vec{k}}{2(2 \pi)^3 k_0}\frac{d^3
  \vec{k'}}{2(2 \pi)^3 k'_0}(2 \pi)^4 \delta^4(p+q-k-r-k')
M^2(p,q,r,k) \\ \times S[k'] \Theta( E_{t1}-(\Emin+\Delta)) \Theta(\Et2
-\Emin) .
\end{multline}

From now on, to ease presentation, 
we shall only consider the calculation for the incoming quark channel 
with partons $r$ and $k$ being respectively an outgoing quark and gluon, exactly as depicted in figure~\ref{fig:Born}. 
For the final results we include all possible configurations, \ie the contribution from the incoming gluon channel as well.

$S[k']$ is the three particle ($qqg$) antenna pattern, which (for our chosen channel) 
is given by:
\begin{equation}
\label{eq:antenna}
S[k']  = g^2 \frac{N_c}{2}
  \left[w_{rk}+w_{pk} -\frac{1}{N_c^2} w_{pr}\right ]\>,
 \end{equation}
where
\begin{equation} 
w_{rk} = \frac{2\>(rk)}{(rk')(k'k)}\>,
\end{equation}
with $N_c =C_A$ , and $g^2$ being the strong coupling such that
$g^2/4 \pi = \alpha_s$.  

At NLO the leading real soft gluon contribution to
the slope $\sigma'$ can be obtained by differentiating \wrt $\Delta$ in
eq.~\eqref{eq:nlo}, which then gives

\begin{multline}
  \label{eq:nlodel}
 C_{1,{\mathrm{real}}}^{'} (\xi,\Delta) \approx -\frac{1}{8 (2 \pi)^5} \int
    \frac{d^2 \vec{r}_{t} d^2 \vec{k}_{t}}{|r_3 k_0 -k_3 r_0|}
    \frac{d^3 k'}{k'_0} M^2 S[k'] \\ \times
    \delta^2(\vec{k}_t+\vec{r}_t+\vec{k'}_t) \Theta(\Et2-\Emin)
    \delta(E_{t1}-\left(\Emin+\Delta \right))
\end{multline}

Using vectorial $k_t$ conservation, we can remove the delta function recalling that it leads to the condition $|k_x| = |E_{t1}-E_{t2}|$. 
From the region $E_{t1} > E_{t2}$ and after introducing the virtual
emission of $k'$ (which has weight $-S[k']$), we obtain the following
expression, valid at small $\Delta$ (see figure~\ref{fig:phasdiag}):
\begin{equation}
  \begin{split}
    C'_1 (\xi,\Delta) &\approx -\frac{1}{8(2\pi)^5}\int \!\!\frac{d^2\vec{r}_t }{|r_3 k_0 -k_3 r_0|}
    M^2 \delta(r_t-(\Emin+\Delta))\\
    &\times\int \frac{d^3 k'}{k'_0} S[k']\left[\Theta(\Delta-|k'_x|) -1\right]
    \>,    
  \end{split}
\end{equation}
where the integration in $k'$ extends to the region where $k'$ is not
recombined with $k$ or $r$ to form a jet.  
Note the step function that restricts $|k'_x|$ in the real piece but not in the negative virtual contribution.

On combining with the
leading (Born order) result and wishing to retain only logarithmic terms in $\Delta$, the slope of the total rate at small
$\Delta$ can be described at NLO as  
\begin{equation}
\label{eq:factor}
C'_{\mathrm{NLO}}(\xi,\Delta) = C'_0+C'_1 = 
C'_{0}(\xi,\Delta)W_{\mathrm{NLO}}(\Delta).
\end{equation}

It is simple to calculate $W_{\mathrm{NLO}}(\Delta)$, although the full answer will depend on the jet-algorithm employed. 
However the leading DL term in $Q/\Delta$ follows by collecting the collinear singularities along the incoming leg $p$ in $S[k']$, assuming an algorithm that ensures that gluons collinear to the final state hard partons 
do not contribute,  and we obtain (recall that we are considering the
incoming quark channel)
\begin{equation}
\label{eq:nlores}
W_{{\mathrm{NLO}}}(\Delta) = 1+\int \frac{d^3 k'}{2(2\pi)^3 k'_0}
S[k']\left[\Theta(\Delta-|k'_x|) -1\right]
\approx 1-C_F \frac{\alpha_s}{\pi} \ln^{2}\frac{Q}{\Delta}\>+\cdots,
\end{equation}
where the dots denote SL terms that we shall compute later, 
and constant pieces 
and terms that vanish at small $\Delta$, which need to be accounted for by matching.

To summarise, at small $\Delta$ one inhibits the radiation of real soft
gluons and this gives rise to the DL behaviour in the
slope $\sigma'$ via the double logarithm in the coefficient function $C'_\mathrm{{NLO}}$.  
This DL
behaviour is responsible for an
unphysical turn over of the NLO calculation at small $\Delta$ \cite{ZEUS}.  
In order
to cure this pathological behaviour one has to perform a resummation
of such soft gluon effects to all orders. 
This will be the aim of the
next section. 
\section{All order result}
The extension of the NLO result eq.~\eqref{eq:factor} to all orders
has two main elements: the computation of multiple all-order soft
gluon emission from a three particle antenna comprising two final
state and one incoming hard parton (global component) and accounting
for correlated emission outside the final state jets from soft gluons
within the jets (non-global logarithms)\cite{DassalNG1,DassalNG2,BMSNG}.  
At NLL level the details of
the jet algorithm become relevant, so that before proceeding with the
calculation it is useful to have a procedure to refer to. 
One can, for example, use the cone
algorithm introduced in~\cite{KidSter}, which samples the phase space for sets of particles flowing into cones of fixed angular size, $\delta$, and 
is defined as follows:
\begin{enumerate}
\item given a set of final state momenta $X=\{q_i\}$, for any subset $x$ of $X$
  compute the unit vector
  \begin{equation}
    \label{eq:unit}
    \vec n_x \equiv \frac{\sum_{i\in x}\vec q_i}{|\sum_{i\in x}\vec q_i|}\>;
  \end{equation}
\item consider the set $x'$ of all particles that flow inside a cone
  of half-angle $\delta$ centered around $\vec n_x$;
\item a jet is any set of particles $x$ for which $x'=x$.
\end{enumerate}
With such a procedure one can, of course, generate any number of jets.
However we are not concerned with the detailed jet structure of the final state, for example identifying the number of jets. We are just concerned with the two highest $E_t$ jets generated by this procedure and the $E_t$ mismatch between them.
We shall consider $\delta$ small compared to 1, 
but not so small that one needs to resum logarithms in the cone-size, $\alpha_s \ln 1/\delta \ll 1$. 
Equipped with this procedure we turn to computing the all-order resummed result. 
We begin by examining the multiple independent emission (global) contribution.

\subsection{Global component}
To derive this part of the result, one just considers that multiple
soft gluons are emitted independently of one-another, each following
the antenna pattern $S[k']$. Secondary parton splitting is built-in via the running coupling. However, in the present case, this approximation misses a subset of the 
next-to--leading logarithms, which arise 
from correlated as opposed to independent 
emission (non-global logarithms)\cite{DassalNG1,DassalNG2}. 
In other words gluons emitted outside but 
near the jet boundaries, feel the dynamical influence of relatively 
harder emissions that are 
inside the jets and this configuration generates next-to--leading logarithms.

The global part of the answer has both leading (double) and
next-to-leading (single) logarithms. The leading logarithms shall
arise from emission soft and collinear to the incoming hard parton.
The next-to--leading logarithms in the global piece will arise from two
sources.  Firstly soft radiation at large angles to the three 
hard emitters is a
source of single logarithms, which will contain a characteristic
dependence on the geometry of the three-pronged hard antenna. An
additional source of single logarithms is hard emissions
quasi-collinear to the incoming parton leg, which we shall also include.

To derive the all orders global result we start with the observation
that the function $W(\Delta)$ can be defined by extending the NLO
result eq.~\eqref{eq:nlores} to all orders, using the factorisation
property of multiple soft gluon ensembles:
\begin{equation}
 W(\Delta)=\frac{1}{n!}\sum_{n=0}^{\infty} \int \prod_{i=1}^{n}\left
      (\frac{d^3 k'_i}{2 (2 \pi)^3 k'_{i0}} S[k'_i]\right
      )\left[\Theta(\Delta-|\sum_{i \notin 1,2} k'_{xi}|) -1\right ]\>,
\end{equation}
where the above integral includes 
contributions only from gluons $i \notin 1,2$, \ie that
fly outside the final state jets $1$ and $2$.
 
We first factorise the step function as follows:
\begin{equation}
  \label{eq:factorise}
  \begin{split}
  \Theta(\Delta-|\sum_{i \notin 1,2} k'_{xi}|)&=
\int_{-\infty}^{\infty} dq_x \Theta(\Delta-|q_x|)\delta(q_x-\sum_{i
    \notin 1,2} k'_{xi})\\
&=\frac{1}{\pi}\int_{-\infty}^{\infty}\frac{db}{b} \sin(b\Delta)
\prod_{i \notin 1,2} e^{ibk'_{xi}}\>,
\end{split}
\end{equation}
where we used the Fourier transform of the delta function 
\begin{equation}
\delta(x) = \int_{-\infty}^{\infty}\frac{db}{2\pi} \exp[-ibx]\>.
\end{equation} 
This allows to simplify the above further to read (valid near $\Delta=0$)
\begin{equation}
\label{eq:C-approx}
C'(\xi,\Delta) = C'_0+C'_1+\cdots = C'_0(\xi,\Delta) W(\Delta)\>,
\end{equation} with 
\begin{equation}
W(\Delta)=\frac{1}{\pi}\int_{-\infty}^{\infty}\frac{db}{b} \sin(b\Delta) 
\exp[- R(b)].
\end{equation}
Here the `radiator' $R(b)$ is given by 
\begin{equation}
\label{eq:rad-def}
  R(b) = \int \frac{d^3 k'}{2(2 \pi)^3 k'_{0}} S[k'] 
  \left ( 1-\exp[ib k'_x]\right )\>. 
\end{equation} 
To NLL accuracy we can simplify the radiator via the replacements
\begin{equation}
  \label{eq:replace}
  1-\exp[ ib k'_x] \to 1-\cos(b k_x') \to \Theta \left (k'_t|\cos\phi| -1/\bar{b} \right ),  \qquad 
  \bar{b} = b\>e^{\gamma_E}\>.  
\end{equation}
This allows us to achieve our final form for $W(\Delta)$, which reads 
\begin{equation}
W(\Delta)=\frac{2}{\pi}\int_{0}^{\infty}\frac{db}{b} \sin(b\Delta) 
\exp[- R(b)].
\end{equation}
with radiator at NLL accuracy given by: 
\begin{equation}
\label{eq:rad}
  R(b) = \int \frac{d^3 k'}{2(2 \pi)^3 k'_{0}} S[k'] 
  \Theta \left (k'_t|\cos\phi| -1/\bar{b} \right )\>.
 \end{equation}
We emphasise that the integral over 
$k'$ is in the region where it is not recombined with an outgoing jet.

We now proceed to the computation of the radiator in the particular
case of an incoming quark. During the calculation we will discuss how
the results obtained can be generalised to the incoming gluon case.  

\subsubsection{Leading logarithmic result}
 At the leading logarithmic (LL) level the radiator is easy to compute
 since one has to consider just radiation collinear to the initial
 state parton. Radiation collinear to either $r$ or $k$ will be
 clustered into a jet and hence in a cone-type or inclusive $k_t$
 algorithm the only source of leading logarithms will be from this
 initial state radiation.
 
 Collecting the
 collinear singularities along the incoming direction in $S[k']$ and
 using eq.~\eqref{eq:rad} we arrive at
\begin{equation}
\label{eq:doublogs}
R_{\mathrm{DL}}(b) = 2 C_F \int \frac{dk'_t}{k'_t} \frac{\alpha_s(k'_t)}{\pi}\; 
\ln \left ( \frac{Q}{k'_t} \right ) \frac{d\phi}{2 \pi}  \ \; 
\Theta(k'_t |\cos \phi| -\bar{b}^{-1}) = C_F \frac{\alpha_s}{\pi} 
\ln^2 \left ( \bar{b}Q\right )+\cdots
\end{equation}
where to extract the double logarithms we froze the coupling at scale
$Q$.

Doing the above integral with the running coupling converts the 
DL contribution into a LL function given by

\begin{equation} 
\label{eq:rad-calc}
R_{\mathrm{LL}}(\bar b)=2 C_F\int_{1/\bar b}^Q \frac{dk'_t}{k'_t}
\frac{\as(k'_t)}{\pi}\ln\frac{Q}{k'_t}=R_{\mathrm{LL}}(b)+\gamma_E
\frac{\partial R_{\mathrm{LL}}(b)}{\partial\ln b}\>,
\end{equation}
where we have performed an expansion of $R_{\mathrm{LL}}(\bar b)$ around $b$, which we are allowed to do since what is left is a subleading contribution.  At NLL accuracy the radiator $R_{\mathrm{LL}}(b)$ has the following expression:
\begin{equation}
  \label{eq:rad-nll}
R_{\mathrm{LL}}(b) = Lg_1(\lambda) +f_2(\lambda)\>,
\end{equation}
where $g_1$ is the leading logarithmic result, $f_2$ is a piece of the NLL contribution $g_2$  and $L \equiv \ln (bQ)$ while $\lambda = \beta_0 \alpha_s(Q) L$.  
The leading logarithmic result $g_1$ reads
\begin{equation}
\label{eq:rad-calcfull}
g_1(\lambda) = -\frac{C_F}{2 \pi\beta_0 \lambda}\left[2\lambda+\ln(1-2\lambda)\right]\>,
\end{equation}
while $f_2$ is given by
\begin{equation}
f_2(\lambda) = \frac{K C_F}{4\pi^2\be_0^2}
\left(\ln(1-2\lambda)+\frac{2\lambda}{1-2\lambda}\right)
             -\frac{\be_1 C_F}{2 \pi \be_0^3}\left(\frac12\ln^2(1-2\lambda)+
\frac{\ln(1-2\lambda)+2\lambda}{1-2\lambda}\right)\>,
\end{equation}
where $\beta_0$ and $\beta_1$ are the first two coefficients of the
QCD beta function:
\begin{equation}
  \label{eq:beta}
  \beta_0=\frac{11C_A - 2\nf}{12\pi}\>,\qquad
  \beta_1=\frac{17 C_A^2-5C_A \nf-3C_F \nf}{24\pi^2}\>.
\end{equation}
In order to account for all NLL contributions coming from soft and
collinear radiation, the coupling constant in the $k_t'$ integral
in~\eqref{eq:rad-calc} has to be taken in the physical CMW
scheme~\cite{CMW}, which is related to the $\overline{{\mathrm{MS}}}$ scheme by the
relation
\begin{equation}
  \label{eq:scheme}
  \as(k)=\alpha_s^{\overline{{\mathrm{MS}}}}(k) \left(1+K
    \frac{\alpha_s^{\overline{{\mathrm{MS}}}}(k)}{2\pi}\right)\>,
  \qquad
  K=C_A\left(\frac{67}{18}-\frac{\pi^2}{6}\right)-\frac{5}{9}\nf\>.
\end{equation}
The logarithmic derivative of $R_\mathrm{LL}(b)$ in~\eqref{eq:rad-calc} can be obtained by differentiating only the $g_1$ piece of $R_\mathrm{LL}(b)$, since what is left is subleading.

Note that although we have labeled the piece of the radiator computed
here as  $R_{\mathrm{LL}}$, we also include in it NLL terms arising
from the running coupling and change of scheme. It is perhaps better
to think of this as a 
DL
piece (arising from soft {\it{and}} collinear emission, while the next-to--leading logarithms we compute subsequently are either pure soft 
or pure collinear SL ($\alpha_s^n L^n$) effects.

\subsubsection{NLL soft contribution}

Having computed the leading logarithmic piece of our answer, which is
independent of the details of the jet definition (\eg the cone size)
we now turn our attention to the NLL terms arising from the
independent emission (global) part of the answer.  Non-global NLL
terms to do with correlated emission will be treated in the next
subsection.  To include global NLL terms we need to compute
eq.~\eqref{eq:rad} using the full form of the soft function $S[k']$,
rather than just collecting the collinear singularities on the
incoming leg, as we did for the leading logarithmic terms.  This will enable us
to correct for SL terms arising from soft, coherent
interjet radiation.  Additionally we have to treat the dependence of
the variable on the azimuthal angle $\phi$, which could be discarded
at leading logarithmic level, and correct for hard collinear emission.

We first treat the $\phi$ dependence.  To NLL accuracy
eq.~\eqref{eq:rad} can be written as (performing a Taylor expansion
about $|\cos\phi|=1$ of the full result and retaining terms only up to
NLL accuracy)
\begin{equation}
\label{eq:radphi}
R \left ( b \right ) = \int \frac{d^3k'}{2(2 \pi)^3 k'_0} S[k] \Theta
\left (k'_t-\frac{1}{\bar{b}} \right )+\frac{\partial  R_{{\mathrm{LL}}}\left (b \right )}{\partial \ln b}  \int_0^{2\pi} \ln (|\cos \phi |) \frac{d\phi}{2\pi}+\cdots
\end{equation}
where the ellipsis denotes terms beyond NLL accuracy which would be
produced by taking higher derivatives or taking the derivative of any
piece of $R(b)$ that is not leading logarithmic. The function
$R_{{\mathrm{LL}}}(b)$ was computed in eq.~\eqref{eq:rad-nll}  
(in fact the only relevant contribution to the derivative above will be from the $g_1$ piece of $R_{\mathrm{LL}}$ and we shall discard other terms)
while $\int_0^{2\pi} \ln (|\cos \phi |) \frac{d\phi}{2\pi}=-\ln2$.

This leaves the first term on the right hand side of the above equation
which contains both the already computed leading-log terms and
next-to--leading logarithms yet to be accounted for. In order to
compute it fully one can treat each dipole in $S[k']$ in turn.  For
example let us consider the $rk$ dipole where $r$ and $k$ initiate the
final state jets. We shall take $r$ and $k$ to be an outgoing quark
and gluon respectively (see figure~\ref{fig:Born}) and the corresponding contribution to the first
term on the right hand side of \eqref{eq:radphi} is
\begin{equation}
\label{eq:raddip}
R_{rk} \left ( b \right ) = N_c\int \frac{d\omega'}{\omega'} \int_{k'\notin \delta_{r,k}}\!\!\frac{d^2\vec{n}_{k'}}{2\pi}\frac{\alpha_s(\kappa'_t)}{2\pi} \frac{[rk]}{[rk'][kk']} \Theta\left ( k'_t-1/\bar{b}\right )
\end{equation}
where $[rk]$ (for instance) denotes $1-\cos\theta_{rk}$, and the scale
$(\kappa'_t)^2 = 2(rk')(kk')/(rk) $ is the transverse momentum
(squared) of the soft emission $k'$ \wrt the emitting dipole pair. This
scale naturally emerges when one considers the collinear splitting of
gluon $k'$ into two offspring partons with similar energies
\cite{BDMZ}.  Note that the angular integration over the directions, 
$\vec{n}_k'$, of
$k'$ in the above equation is constrained such that $k'$ is outside a
cone of angular size $\delta$ around the outgoing hard partons $r$ and
$k$, $k' \notin \delta_{r,k}$. This is of course an approximation since, 
in the chosen algorithm,  $\delta$ is really the allowed 
opening angle wrt the energy weighted centroid of the outgoing hard parton and the emitted soft parton. The correction terms are proportional to powers of the gluon energy and we are entitled to neglect them here, for the small $\Delta$ behaviour.

In principle the integral in eq.~\eqref{eq:raddip} is rather
cumbersome to evaluate in full detail. However one can simplify the
situation to extract the LL and NLL dependence. In particular the $rk$
dipole does not contribute any leading logarithms since these arise
from emissions that are soft and collinear to the incoming leg $p$.
Hence the contribution from the $rk$ dipole is at most NLL in $b$
(recall that we do not attempt to resum logarithms in the cone size
$\delta$). The single logarithms in question arise from the pole in
the integration over energy $\omega'$ and are a soft wide-angle
contribution.  To extract this piece we can simplify \eqref{eq:raddip}
to give
\begin{equation}
\label{eq:smallcone}
R_{rk}(b) \approx C_A \int \frac{d\omega'}{\omega'} \frac{\alpha_s(\omega')}{\pi} \Theta\left (\omega'-1/b \right ) \ln \frac{2[rk]}{\delta^2}\>,
\end{equation}
where we set $N_c = C_A$.  Notice that we felt free to mistreat the
argument of $\alpha_s$ but retained its essential dependence on the
energy $\omega'$ in doing which we neglected a constant of
proportionality which would produce only NNLL terms beyond our
accuracy. In writing the above we also made use of the result
\begin{equation}
\int_{k'\notin \delta_{r,k}}\frac{d^2\vec{n}_{k'}}{2\pi} \frac{[rk]}{[rk'][kk']} = 2 \ln \left ( \frac{[rk]}{1-\cos\delta} -1 \right ) = 2 \ln \left ( \frac{2[rk]}{\delta^2} \right )+\mathcal{O} (\delta^2/[rk])\>,
\end{equation}
and discarded terms involving the ratio of the cone-size (squared) to the interjet separation, which we shall do
throughout this paper. However note that we can retain the full dependence on
cone size, in this wide-angle global piece, by employing the exact
formula mentioned above rather than retaining simply the logarithmic
dependence on cone-size $\delta$. Note also the dependence on the geometry
$[rk]$ of the underlying dipole emitters, that is typical of soft
interjet radiation~\cite{BDMZ}.

Similarly one can compute the other dipoles $pr$ and $pk$. Here we
will also encounter leading logarithms from when $k'$ is near the
incoming leg $p$, and in this region the argument of the running
coupling will reduce to $k'_t$ rather $\omega'$. However we have
already computed these leading/double logarithms in \eqref{eq:rad-calc} and
the remaining NLL piece of the answer will once again be obtained by
arguments along the lines above.  Assembling the contribution from all
dipoles we have below the full soft contribution to the radiator which
can be expressed as
\begin{equation}
\int \frac{d^3k}{2(2 \pi)^3 k_0} S[k] \Theta \left (k_t-\frac{1}{\bar{b}} \right ) = R_{{\mathrm{LL}}}\left (\bar{b} \right )+ R_\mathrm{{NLL}}(b)\>,
\end{equation}
with $R_{\mathrm{LL}}$ as given in eq.~\eqref{eq:rad-calc}
and  $R_{\mathrm{NLL}}$ being the soft global NLL contribution to the
radiator:
\begin{equation}
\label{eq:single-logs}
R_{\mathrm{NLL}} = \left [2 C_F \ln \left (\frac{2[pr]}{\delta} \right
  )+C_A \ln \left (\frac{2[kr][pk]}{\delta^2 [pr]}\right ) 
 + C_F
    \ln\frac{E_p^2}{Q^2}
\right ] (2 t).
\end{equation}
 We introduced above the SL evolution variable $t$:
\begin{equation}
\label{eq:tdef}
t \equiv \int_{1/b}^{Q} \frac{d\omega}{\omega}\frac{\alpha_s(\omega)}{2\pi} = \frac{1}{4\pi \beta_0} \ln \frac{1}{1-2\lambda}.
\end{equation}
Recall that we treated $p$ and $r$ as quarks and $k$ as a gluon, but for the final results we sum over all configurations with appropriate modifications to the above form.

Notice that the SL contribution~\eqref{eq:single-logs} depends not
only on the geometry of the three-parton antenna (specifically on the
angles between hard emitting partons) and the cone size $\delta$,  but
also on the incoming parton energy $E_p$. The additional term $C_F\ln
E_p^2/Q^2$ accounts for the fact that a soft gluon collinear to $p$ has energy up to $E_p = Q/(2\xi)$, and not $Q$, as one would infer from~\eqref{eq:rad-calc}. 
 
There are still two sources of single logarithms missing from the
above answer.  The
first source of single logarithms is from non-soft emissions almost
collinear to the incoming parton $p$, which we shall include next, to complete the global piece of the calculation. 
The other piece we will need is the non-global term arising from 
soft correlated emission of $k'$ from
gluons included within the jets. 

\subsubsection{NLL terms from hard collinear emissions}
Here we note that multiple hard emissions on the incoming leg $p$ also contribute a class of single logarithms, precisely as in the case of DIS event shape variables~\cite{ADS,Dassalbroad,BMSZ}. To exponentiate this piece one has to turn to Mellin space \wrt the Bjorken $x$ variable. 
However we shall directly note (referring the interested reader to the manipulations described for instance in~\cite{ADS}) 
that restricting the $k_t$ of hard 
collinear emissions on the incoming leg, one essentially restricts the DGLAP evolution of the structure function to the scale $1/b^2$, rather than $Q^2$, 
$f(x/\xi,Q^2) \to f(x/\xi,1/b^2)$. 
One also needs to change the scale in the calculation for $R_\mathrm{{LL}}$ such that the corresponding virtual corrections are properly treated. This leads to the replacement of $Q$ in the integrand of eq.~\eqref{eq:rad-calc} by the factor $Q\>e^{-3/4}$. Thus an additional term appears in the radiator which is given by
\begin{equation}
\label{eq:rad-hc}
R^{\mathrm{hc}}_{\mathrm{NLL}} = -\frac{3}{4}\frac{C_F}{\pi\beta_0} \ln \frac{1}{1-2\lambda},
\end{equation}
and the remaining piece of the hard collinear emission is embodied in
a change of scale of the parton densities as mentioned before.
In the incoming gluon case the hard collinear contribution can be
obtained from~\eqref{eq:rad-hc} by simply replacing $-3/4$ with
$-\pi\beta_0/C_A$, with $\beta_0$ defined in~\eqref{eq:beta}.

\subsection{non-global component}
So far we accounted only for independent multiple soft gluon emission, with corrections for hard-collinear emission. 
For a class of observables that typically involve 
angular cuts in the phase space, the independent emission approximation is not sufficient to generate the full single logarithms~\cite{DassalNG1,DassalNG2}. 
Our observable is indeed such a non-global observable, due to the fact that it is 
sensitive to radiation only outside the jets. Hence a soft emission near the jet boundary, which contributes to our observable, can resolve relatively harder emission inside the jets, \ie the jet structure, at NLL level.

We now account for the non-global piece of the final result, for the
slope in $\Delta$ of the dijet rate $\sigma(\Delta)$.  As we said, non-global
single logarithms, $\alpha_s^n \ln^{n}(bQ)$, arise from soft emissions that
fly inside the jets defined by cones, which themselves emit outside
the jet region.  On a heuristic level, emission from jets at large
angles compared to the angular extent of the jets themselves, will see
only the total colour charge of the system of partonic emitters that
constitute the jet.  Thus it follows from coherence properties of QCD
radiation that when one considers the relatively small cone
approximation (jet cones significantly smaller than the interjet
separation) the non-global component will arise separately from each
cone boundary.  We illustrate our ideas by first performing below an
analytical computation of the leading $\alpha_s^2 \ln^2{bQ}$ non
global piece and follow that by considering non-global effects at all
orders.

Before we move on we should however mention that non global logs are rather sensitive to the exact details of the jet algorithm employed. For instance in our case (cone algorithm), a situation could arise 
where a soft parton may form a jet with the hardest (high $E_t$) parton and also with a softer parton itself outside the high $E_t$ jet. The decision on how to attribute the common energy between the jets will affect the size of the non global component. In what follows we ignore this complication and stick to our previous definition of the high $E_t$ dijet, which will mean we take a scenario where non-global logs make a maximal impact.

In order to proceed we need to consider the emission of two soft gluons by a hard three-particle antenna. This has a colour dipole structure identical to that
of the single emission term eq.~\eqref{eq:antenna} except that for
each dipole term $w_{ab}$ of eq.~\eqref{eq:antenna} one inserts the
result for emission of a soft two-parton system by the dipole $ab$,
$w_{ab}^{(2)}$, that is also the relevant function in the two-jet
case (see~\cite{DMO,BDMZ,CatGraz}).  In fact for the non-global term
we shall need to consider only a specific piece of the correlated
two-parton emission term, corresponding to the situation when the two
soft emitted partons are energy ordered, $\omega_1 \gg \omega_2$. Its
detailed form will be mentioned below.

We thus again consider a generic dipole $ab$ formed by two of the
three hard partons that are present in the underlying hard event. We
parametrise the four-momenta of these hard partons as below, along
with those of the emitted two soft gluons $k_1$ and $k_2$:
\begin{equation}
  \label{eq:papbk1k2}
  \begin{split}
    p_a &= E_a (1,0,0,1)\>,\\
    p_b &= E_b (1,0,\sin\theta_{ab}, \cos\theta_{ab})\>,\\
    k_1 &= \omega_1 (1,\sin\theta_1 \sin\phi_1, \sin\theta_1\cos\phi_1,
    \cos\theta_1)\>,\\
    k_2 &= \omega_2 (1,\sin\theta_2 \sin\phi_2, \sin\theta_2\cos\phi_2,
    \cos\theta_2)\>,
  \end{split}
\end{equation}
and assume strong energy ordering $\omega_1 \gg \omega_2$ as is
required to generate the non-global piece.  To trigger the non-global
contribution, the harder parton $k_1$ flies inside the jet cones while
$k_2$ is outside. For the dipole $ab$ we assume that parton $b$ is the
hard incoming parton $p$ while $a$ is an outgoing hard parton ($r$ or
$k$), which initiates a jet with angular size $\delta$. In general
both legs of the dipole $ab$ can correspond to the two outgoing jets
(\eg $a=r,b=k$) and this contribution will be treated subsequently.
For the present case however non-global logs are generated by the
configuration
\begin{equation}
\label{eq:angles}
\cos\theta_1 \geq \cos \delta \, , \, \cos\theta_2 \leq \cos \delta. 
\end{equation}

The leading non-global contribution is then obtained similarly 
as in~\cite{DassalNG1,DassalNG2} by considering the integral 
(once again we only need to consider the dependence on the energy $\omega$, since this is a pure soft piece)
\begin{equation}
-2\left( \frac{\alpha_s}{2\pi}\right)^2\int_0^1\frac{d\omega_1}{\omega_1}\int_0^{\omega_1}\frac{d\omega_2}{\omega_2}\frac{d^2\vec{n_1}}{2\pi}\frac{d^2\vec{n_2}}{2\pi} \, w^{(2)}_{[ab]}  \, \Theta \left (\omega_2 - 1/b \right)\>,
\end{equation}
and  we denoted by $w^{(2)}_{[ab]}$ the angular dependence of $w^{(2)}_{ab}$.
Performing the energy integrals is trivial and gives 
$-\left ( \frac{\alpha_s}{2\pi} \right )^2 \ln^2(bQ)$.
To work out the coefficient of this SL term, we need to integrate  $w^{(2)}_{[ab]}$ over the allowed directions of $k_1$ and $k_2$, $\vec{n_1}$ and $\vec{n}_2$.
For this we need just the angular dependence of the piece of the 
correlated emission term $w^{(2)}_{ab}$which is given by~\cite{DMO}
\begin{equation}
  \label{eq:wab}
  w^{(2)}_{[ab]} = C_A \left (\frac{[ab]}{[a1][12][2b]}+\frac{[ab]}{[a2][21][2b]}-
  \frac{[ab]^2}{[a1][1b][a2][2b]}\right)\>,
\end{equation}
with, as before, $[ij]=1-\cos\theta_{ij}$. 

We first perform an
azimuthal average of $w^{(2)}_{[ab]}$ and get\footnote{We are free to do
  this since one can discard at the SL level the observable's
  dependence on azimuth and just consider its energy dependence as
  indicated above.}
\begin{equation}
  \label{eq:wab-average}
  \langle w^{(2)}_{[ab]} \rangle = \frac{2 C_A\>(1-\cos\theta_{ab})
    \>\Theta(\cos\theta_2-\cos\theta_{ab})}
  {(1-\cos\theta_2)(\cos\theta_1-\cos\theta_2)(\cos\theta_1-\cos\theta_{ab})}
  \>.
\end{equation}
In doing the calculation we have exploited the fact that the only
contribution to non-global logs arises when the harder 
gluon $k_1$ is emitted
inside the jet-cone around $a$ and the softer gluon $k_2$ is emitted
outside. Note that both gluons are soft compared to the hard scale $Q$.

Now we need to integrate over polar angles, which gives
\begin{equation}
 \label{eq:F}
  \int_{\cos\theta_{ab}}^{\cos\delta}\frac{d\cos\theta_2}
  {1-\cos\theta_2}\int_{\cos\delta}^1d\cos\theta_1
  \frac{2 C_A(1-\cos\theta_{ab})}{(\cos\theta_1-\cos\theta_2) 
    (\cos\theta_1-\cos\theta_{ab})}\> = C_A \frac{\pi^2}{3},
\end{equation}
independent of the cone size $\delta$.  The fact that the result does
not depend on the cone size illustrates the fact that the contribution
from emission of $k_2$ is an edge effect arising from the vicinity of
the cone boundary and the geometry of the interdipole region becomes
unimportant, since it corresponds essentially to an infinite interval
in rapidity between the cone boundary and the other (incoming) emitter
(see~\cite{DassalNG2}).

This is no longer true when one considers the dipole formed by the two
outgoing partons and in this case there are contributions from each
cone boundary. The final result depends both on the cone-size
$\delta$ as well as the interdipole separation $[ab]$. However if one
considers the relatively small cone limit such that one neglects terms 
that vary as $\delta^2/[ab] \ll 1$, the
result from such a dipole is simply twice that in \eqref{eq:F}.
Recall that such finite cone-size effects also arose as corrections to
the $\ln 1/\delta$ piece of the global single logarithms and we choose
to neglect them.

The final result for the leading non-global term produced by the
three-jet system then is given by adding up the contributions from
each of the three hard emitting dipoles with the appropriate colour
factors ($N_c/2$ for a quark-gluon dipole and $- 1/2N_c $ for a
quark-antiquark dipole).  
Denoting the entire non-global contribution by the series 
\begin{equation}
S(t) = \sum_{n=2} S_n t^n,
\end{equation}
with $t$ the SL evolution variable introduced earlier in eq.~\eqref{eq:tdef}, we have computed the first term $S_2$ which reads
\begin{equation}
S_2 = -C_A (C_1+C_2) \frac{\pi^2}{3}.
\end{equation}
Here $C_1$ and $C_2$ are the charges of the partons that initiate the
outgoing jets in a given underlying hard configuration.
For our chosen channel with incoming quark, we have $C_1+C_2 = C_F+C_A$.
The fixed order result above is correct up to terms $\delta^2/[rk]$ where $r$ and $k$ are the outgoing hard partons.

To generalise the above result to all orders one has to consider
configurations of several wide-angle soft gluons inside the jet cones,
that coherently emit a single softest gluon outside the jets.  
At present this can only be done in the large $N_c$ limit which reduces the problem to planar graphs, and the corrections to these are suppressed by 1/$N_c^2$. Therefore one expects these neglected terms 
to make a difference at the $10\%$ level to our non-global result 
\cite{DassalNG1,DassalNG2,BMSNG}. 

In this limit one can consider that the contribution of each initiating dipole 
will be modified by a non-global factor\footnote{Strictly speaking this will be true of only the quark-gluon dipoles which survive the large $N_c$ limit. However doing the same also for the large $N_c$ suppressed 
quark-quark dipole will ensure compatibility 
with the leading order result. Any differences with the correct full 
answer will of course be suppressed as $1/N_c^2$.}
\begin{equation}
\label{eq:ngdip}
e^{-R_{ab}} \to e^{-R_{ab}}S_{ab}
\end{equation}
where $e^{-R_{ab}}$ is just the independent emission radiator computed
earlier, for dipole $ab$ and $S_{ab}$ is the accompanying non-global
factor (see arguments in \eg~\cite{BMSNG}).  For dijets with opening
angles $\delta$ that are small compared to the interdipole opening
angle $[ab]$, the contribution $S_{ab}$ coming from each cone boundary will
in fact be the same as that obtained for the case of a emission into a
large (effectively infinite) rapidity slice, already derived for the
two-jet (single initiating dipole) case~\cite{DassalNG2}.
This property shows up in the fixed order computation but it was also 
formally shown at all-orders in~\cite{BMSNG} that for problems with limited energy flow everywhere except in small cones around the leading hard partons, the non-global contribution arises separately from each cone boundary. This ensures that one can extend the results derived previously~\cite{DassalNG1,DassalNG2} for emission into an infinite rapidity region, to the present case. The corrections to this result will vary as the square of the cone-size $\delta$.

Putting together the contribution of all the three initiating dipoles
(the all order non-global result is computed in the large $N_c$
limit) corrects the SL global result described
earlier:
\begin{equation}
e^{-R(b)} \to e^{-R(b)}S(b) \>,
\end{equation}
where $S(b)$ can be parametrised as below~\cite{DassalNG1}:
\begin{equation}
  \label{eq:MainResult}
  S \simeq \exp\left( - (C_1+C_2) C_A \frac{\pi^2}{3}
     \left( \frac{1 + (at)^2}{1 + (bt)^c}\right)t^2\right),
\end{equation}
with $t$ as in \eqref{eq:tdef} and 
\begin{equation}
  a = 0.85C_A\,,\qquad b = 0.86 C_A\,, \qquad c = 1.33\,.
\end{equation}

We have retained through the above parametrisation 
the correct $C_A(C_1+C_2)$ colour structure for the leading term $S_2$ 
calculated earlier, but beyond this leading term the result is correct only in the large $N_c$ limit.
The paramtrisation above is valid for $t \leq 0.7$, which is more
than sufficient for our purposes. This is because, in practice, 
the $b$ integral
receives vanishingly small contributions near and beyond this point,
which corresponds to $\beta_0 \alpha_s L = 0.497$, very close to the
Landau pole value of $0.5$.
In practice however, non-global logs make a very small contribution to
the overall result. This is due to the fact that they start at
$\mathcal{O}(\alpha_s^2)$, relative to the Born term, and in the
region where they may be expected to formally be important (at very
small $\Delta$), the $b$ space integral is dominated by the 
DL
behaviour.

\section{Results and general properties}
\label{sec:general}
Here we shall present our final result and illustrate its main properties. 
We can express our resummed result in the form (valid at small $\Delta$)
\begin{equation}
  \label{eq:slope-def}
    \sigma'(\Delta)\approx \int_x^1 \frac{d\xi}{\xi}C'_0(\xi,\Delta)W\left(\Delta,\frac{x}{\xi}\right)\>,
\end{equation}
where we suppressed the dependence on $Q$ and $E_{\mathrm{min}}$ and $C'_0$ 
is the Born level coefficient function for the slope defined in 
eq.~\eqref{eq:slope-Born}.
We redefined the all orders quantity $W(\Delta)$, to include the
dependence on the parton densities,
\begin{equation}
\label{eq:W-eval}
W \left (\Delta,\frac{x}{\xi}\right ) = \frac{2}{\pi} \int_0^{\infty} 
\frac{db}{b} 
\sin(b \Delta) e^{-R(b)}S(b)f\left(\frac{x}{\xi},\frac{1}{b^2}\right).
\end{equation}

Here $R(b) = R_{\mathrm{LL}}(\bar b)+R_{\mathrm{NLL}}(b) +R^{\mathrm{hc}}_{\mathrm{NLL}}(b),$ is the radiator computed in the text, in terms of the separate pieces indicated.
$S(b)$ is the non-global contribution. 

The most important feature of our answer is the absence of a Sudakov peak when one goes from $b$ space to $\Delta$ space.
This feature emerges at the
DL
level itself, and hence while SL
effects make a significant numerical difference to the final answers,
the properties pointed out in the following discussion are essentially
unaffected by neglecting the single logarithms.
Therefore for this discussion, the relevant quantity we examine is 
\begin{equation}
\label{eq:basic}
W_\mathrm{DL}(\Delta) = \frac{2}{\pi}\int_0^\infty \frac{db}{ b} \sin(b \Delta) e^{-R_\mathrm{DL}(b)}\>,
\end{equation}
where DL indicates that we kept only the DL terms in $b$ space and use for $R_{\mathrm{DL}}$ the expression given in eq.~\eqref{eq:doublogs}.

As was pointed out in Ref.~\cite{azimcorr} the integral above has two
distinct regimes. In order to study these separately we divide the
integral above as follows
\begin{equation}
\int_0^\infty db = \int_0^{b_0} db+ \int_{b_0}^{\infty} db\>,
\end{equation} 
where $b_0$ is taken as $1/\bar{\Delta}$ and $\bar{\Delta} = \Delta
e^{\gamma_E}$. The discussion that follows holds also for other
possible choices of  $b_0$, as long as its value is of order $1/\Delta$. 

First we shall deal with the second term on the right-hand side of the above
equation, where the integral is dominated by its lower bound $b_0$.
For this term one can invert the $b$ transform retaining only terms
that will contribute up to single logarithmic accuracy and write
\cite{azimcorr}
\begin{equation}
\begin{split}
\frac{2}{\pi}\int_{1/\bar{\Delta}}^{\infty}\frac{db}{b}\sin(b\Delta)
\,e^{-R_{\rm DL}(b)}
\approx e^{-R_{\rm DL}(1/\bar{\Delta})}
\left[\frac{e^{-\gamma_E R'}\sec\frac{\pi}{2}R'}{\Gamma(1+R')}-
\frac{2}{\pi}\sum_{n=0}^\infty\frac{(-1)^n}{(2n+1)!}
\frac{e^{-(2n+1)\gamma_E}}{2n+1-R'}\right]\>,
\end{split}
\end{equation}
where the function $R'$ is
\begin{equation}
R'(1/\Delta) = \frac{2\as C_F}{\pi}\ln\frac{Q}{\Delta}\>.
\end{equation}
The result above corresponds to a Sudakov behaviour with a peak at $R' = 1$.

However we have only considered the contribution from $b > b_0$. The
reason for doing this is that attempting to describe the whole $b$
integral by a Sudakov behaviour (and retaining only terms up to SL accuracy as above) would produce only the first term on
the right-hand side of the above equation, which is divergent at $R'=1$. This is
an indication that near this region, a Sudakov
behaviour is not the dominant contribution.

The physical reason for the above statement is mentioned below and appears in
many other examples including the Drell-Yan $p_t$ distribution~\cite{ParisiPetronzio,RakowWebber}.
At
very low $\Delta$ (below the peak region, $R' \approx 1$) the dominant
mechanism by which one obtains a low $E_t$ difference between final
state jets is vectorial cancellation, rather
than soft emission. In fact this mechanism is dominant at small
$\Delta$ since only a one-dimensional cancellation is required,
$|\sum_i k_{xi}| \to 0$, in the present case. Hence the Sudakov peak
at $R'=1$ is washed out. 
In problems requiring a two-dimensional vectorial cancellation~\cite{ParisiPetronzio}, the vectorial cancellation is important only beyond the Sudakov peak which therefore appears.

To see how the recoil cancellation mechanism behaves, we have to
consider the small $b$ contribution :
\begin{equation}
\label{eq:smallDelta}
\begin{split}
\frac{2}{\pi}\int_0^{1/\bar\Delta}
\frac{db}{b}\>
\sin(b\Delta)
e^{-R_\mathrm{DL}(b)}
= \frac{2}{\pi}\sum_{n=0}^\infty
\frac{(-1)^n(\Delta)^{2n+1}}{(2n+1)!}
\int_0^{1/\bar\Delta}db\,b^{2n}
e^{-R_\mathrm{DL}(b)}\>,
\end{split}
\end{equation}
where we expanded the sine function. In general, this integral has to
be done numerically. However, the simple form of $R_\mathrm{DL}(b)$
allows us to integrate eq.~\eqref{eq:smallDelta} term-by-term analytically,
and obtain
\begin{equation}
\int_0^{1/\bar\Delta}db\,b^{2n}
e^{-R_\mathrm{DL}(b)}=
\frac{e^{-(2n+1)\gamma_E}}{Q^{2n+1}}\frac{e^{\frac{(2n+1)^2}{2a}}}{\sqrt{a}}
\Phi\left(\sqrt{a}\ln\frac{Q}{\Delta}-\frac{2n+1}{\sqrt{a}}\right)\>,
\end{equation}
where the quantity $a$ and the function $\Phi(x)$ are
\begin{equation}
a = \frac{2\as C_F}{\pi}\;,\qquad
\Phi(x) = \int_{-\infty}^x dt\, e^{-\frac{1}{2}t^2}.
\end{equation}
 
The series above is rapidly convergent. At very small $\Delta$ one
obtains the leading behaviour
\begin{equation}
W_{\mathrm{DL}}(\Delta) \approx \left ( \sqrt{\frac{2}{\pi a}}
  e^{-\gamma_E}e^{1/2a} \right )\frac{\Delta}{Q}\>,
\end{equation}
which means that the slope of the resummed dijet distribution near $\Delta =0$
goes linearly to zero, with a proportionality constant which behaves as
$1/\sqrt{\alpha_s}$ which arises from integrating a Sudakov form factor.
This behaviour may of course receive corrections by terms that start at relative order $\alpha_s$, and we need to account for these pieces with a matching procedure.

The actual value of the coefficient of the linear behaviour 
at small $\Delta$ will obviously also depend on the SL
 terms, but the qualitative behaviour 
is as we have demonstrated at the DL level.

We next wish to 
present some plots for the resummed slope $\sigma'(\Delta)$ for 
dijet
production in the Breit frame of DIS.
However before doing that 
we make some additional points about the numerical 
computation. 
In order to ensure physical behaviour over the entire range of $b$ integration, we can redefine the resummation variable $bQ$, by making the replacement 
$\Sigma(b)\to
\Sigma(Q^{-1}\sqrt{1+b^2 Q^2})$~\cite{Higgs}.
This is an ad-hoc prescription to some extent 
but 
it leaves the soft region (large $b$ behaviour of the radiator) unaltered, and does not change the linear behaviour at small $\Delta$, that we described earlier.
Other such prescriptions are also possible but would differ only in the large $\Delta$ region, which in any case requires matching to fixed order.
We also point out that to avoid the Landau pole in the running coupling 
(and unphysical behaviour of the parton distributions) we have to put a cut-off on the $b$ integral at large $b$. We take this to be at the Landau pole, $\lambda =\beta_0 \alpha_s \ln bQ =0.5$, although varying the position of this cut near the vicinity of this value makes no noticeable difference at all to the results. We shall comment on the role of non-perturbative effects in our conclusions.

In all the presented plots we choose for the DIS variables the values
$x=0.01$ and $Q=20\,\mbox{GeV}$, consider dijet events with the cone algorithm
described earlier and fix the opening angle
$\delta=0.2$ with the minimum transverse energy $\Em=10 \,\mbox{GeV}$. In
order to have two well separated jets which respect the condition 
$\delta^2/[ij] \ll 1$ , where $[ij]$ is the interjet opening angle, 
we impose a cut $|\eta| \leq 1$ on the
rapidity $\eta$ of the jets (with respect to the photon direction) 
in the Breit frame. 
This works in our case, since for the basic Born configuration we 
integrate over, the jets must have equal (cancelling) 
transverse momenta \wrt the photon direction. On an experimental level (or beyond leading order jet production), the cut \wrt the photon axis 
is not a sufficient means to ensure the separation of the high $E_t$ dijets, and we need to impose a cut on say the interjet rapidity. 
This is however not needed for our purposes here. Additionally we use the 
MRST2001\_1 parton data set~\cite{MRST} and the pdf 
evolution code described in 
Ref.~\cite{Dassalbroad}.

In figure~\ref{fig:ratio} we plot the ratio
\begin{equation}
  \label{eq:ratio}
  D(\Delta) =  \sigma'(\Delta)/\sigma'_0(\Delta)\>,
\end{equation}
where $\sigma'_0(\Delta)$ is the Born value for the
slope~\eqref{eq:slope-Born}.  It is interesting to compare the
behaviour obtained after resummation  (the upper curve), with the
logarithmically enhanced  part of the NLO result, which we have
computed.
\FIGURE { \epsfig{file=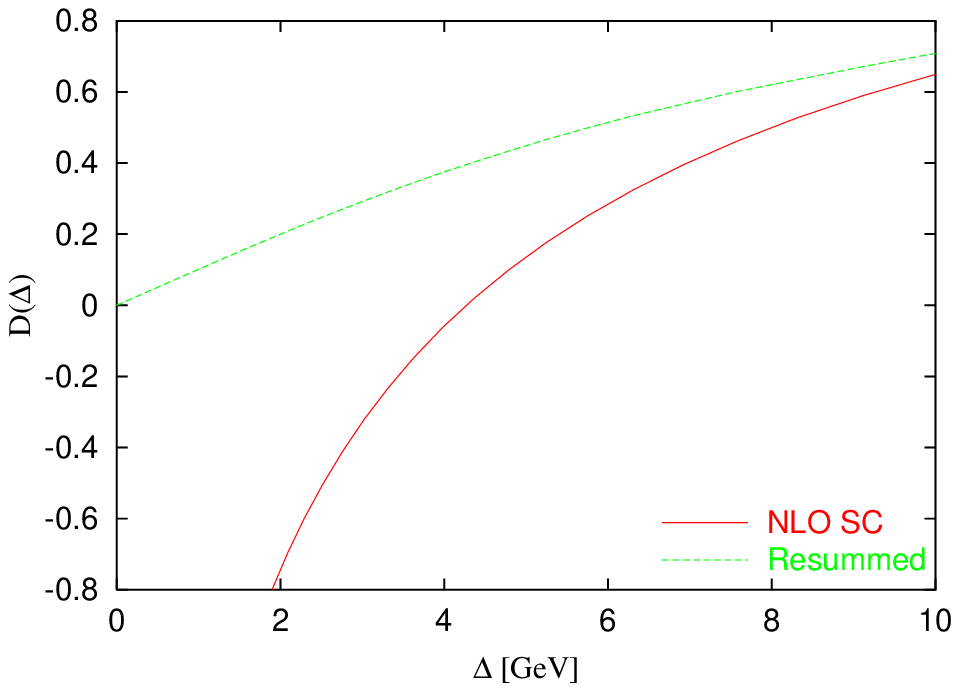, width=.6\textwidth}
  \caption{The ratio $D(\Delta)$ at LL and NLO (soft/collinear) accuracy, for $Q=20 \, \mbox{GeV}$ and $x=0.01$.}
  \label{fig:ratio}}
 While the resummed result has the same sign as the Born term (both are negative as required on physical grounds), the NLO result takes the opposite sign to the Born term, at small $\Delta$, and becomes divergent.

Note also that the resummation predicts a slope that 
vanishes linearly at small $\Delta$, while the leading (Born) order slope is finite in this limit. The corrections to this resummed result will, at small $\Delta$, at best 
start at relative 
order $\alpha_s$. Hence we still (even after matching to NLO) 
expect a value of the slope that in the small $\Delta$ region, 
is much smaller than that obtained at leading order.

Although we do not explicitly indicate it, 
it should be understood that the integration over the 
hard Born configuration is done with the angular cuts described above. The 
final results include also the gluon incoming channel as well as the transverse and longitudinal contributions to the result. 

In figure~\ref{fig:llvsnll} we can see the impact that NL logarithms
have on the slope. There we compare the full NLL answer
\eqref{eq:W-eval} with the LL result, obtained from \eqref{eq:W-eval}
by neglecting all NLL effects, \ie
setting $R(b)=R_\mathrm{LL}(b)$, freezing the scale of the pdf's at
$Q$ and making the replacement $S(b)\to 1$.  One can observe that both curves show a linear behaviour for small $\Delta$ but that the LL curve is steeper than the NLL one. The non-global component of the NLL result however,  makes little difference to the final results (actually negligible in the very small $\Delta$ region and less than $5\%$ in the whole selected range). The reasons for this were mentioned before and this observation is perhaps consoling in light of the fact that these are only known in the large $N_c$ limit.  \FIGURE { \epsfig{file=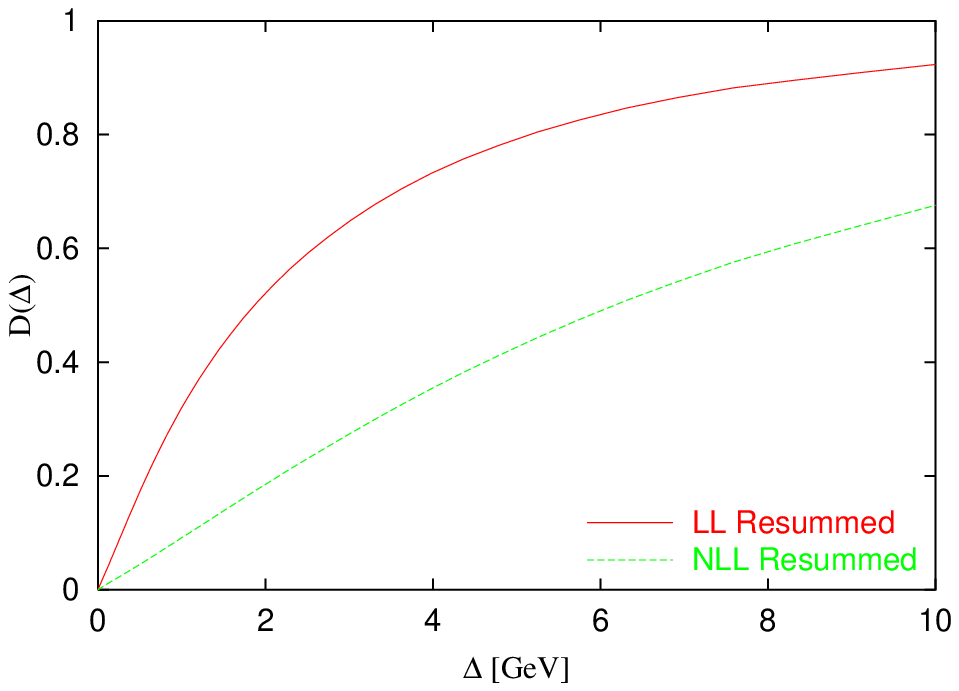, width=.6\textwidth}
  \caption{The ratio $D(\Delta)$ at LL and NLL accuracy, for $Q=20 \, \mbox{GeV}$ and $x=0.01$.}
  \label{fig:llvsnll}
  }




\section{Conclusions}
Here we present the main conclusions from this work. 
We have demonstrated that the region of symmetric $E_t$ cuts in dijet production can be handled with an all order resummation. 
%
The resummation is carried out in the slope of the dijet rate $\sigma'(\Delta)$.  The result we obtain shows that the slope of the resummed 
dijet rate $\sigma'(\Delta)$ is negative throughout and vanishes at
small $\Delta$, at least up to corrections of relative order $\alpha_s$. The NLO calculation on the other hand is positive and divergent at small $\Delta$.

The resummation we carried out enables us to probe the $E_t$ distribution of the higher energy jet, arbitrarily close to the cut on its energy, which would otherwise be unsafe. 
Additionally after matching to fixed order, one can in principle achieve a better estimate of the rate at $\Delta =0$, $\sigma(\Delta)-\sigma(0) = \int_0^\Delta dx \sigma'(x)$, where one can choose $\Delta$ in the region where the NLO calculation is reliable and hence use the NLO value. In fact choosing $\Delta$ large enough that the rate $\sigma(\Delta)$ vanishes 
would enable the direct determination of $\sigma(0)$.

Our calculation was performed using a simple variant of the cone algorithm~\cite{KidSter}. As we stressed in the paper, other variants of this algorithm can be employed and would lead to different results arising from NLL differences. 
However we still expect our result to serve as a qualitative 
model for the NLL terms, in all cases where the 
recombination scheme ensures that the $E_t$ mismatch between the dijets is given by the component $k_x$ (defined earlier) of soft gluon momenta flowing outside the jets.
This is because however the jets are defined, 
soft radiation at large angles to the jets will follow a
three-particle antenna pattern identical to the one employed here. The
terms that depend on the geometrical details of the algorithm, \ie the finite cone size pieces, will 
of course be different but computable at least in the global term. 
For algorithms that involve clustering such as the inclusive $k_t$ algorithm with an $R$ parameter, the situation with non-global logs will be different. 
In fact one expects the clustering procedure to reduce this component~\cite{AppSey}, which in any case does not significantly affect the result, and hence 
we should be able to extend our computation to apply to that case.

We have not mentioned, thus far, that in general one may expect a small $E_t$ 
to be generated on the incoming leg, due to non-perturbative effects (intrinsic $k_t$). 
This would also lead to a mismatch between the jet $E_t$'s. However this effect is expected to correct the radiator at the level of terms quadratic in $b$, which would lead to a $1/Q^2$ power correction, as in the case of the Drell-Yan $p_t$ distribution~\cite{NPDY}.

Another important development that is needed is the matching to fixed order of our calculation. This would ensure that one can describe the slope at the two-loop level completely (accurate even at larger $\Delta$), 
if one replaces the pathological 
logarithmic behaviour with our all-order resummation.

We also mention that it is possible to perform the resummation 
in other ways, to obtain a finite rate with symmetric cuts.
For instance one can resum threshold logarithms in the invariant 
mass distribution of the dijet pair. Then one can probe the mass distribution safely, even in the highest mass bin, with symmetric cuts. An integration over all mass bins would yield a rate that is finite with symmetric cuts. This will be the procedure adopted in~\cite{BanDasprep}.

As further extensions of this work, one can conceive of dijet production via the resolved photon contribution rather than DIS. This would involve a four 
particle antenna, rather than one with three particles, as was the case here.
Similar issues will also arise in the case of dijets produced via $\gamma \gamma$ collisions at LEP~\cite{LEP} and the case of prompt di-photon hadroproduction as we mentioned before~\cite{DelDuca}.

Lastly we hope that the work carried out here will eventually lead to 
comparisons with experimental data. However much work remains 
to be done in terms of matching to fixed order, which will include having to consider jet production beyond leading order, as well as 
adjusting the details of the calculation to a different jet algorithm. 
We intend to address these issues in 
forthcoming work.

\acknowledgments{
We wish to thank the following people for useful discussions: Stefano Catani, 
Stefano Frixione, Guenter Grindhammer, Roman Poeschel, Gavin Salam and Mike Seymour. Additionally we would like to thank Giulia Zanderighi for a careful reading and comments on the manuscript.
Lastly we also wish to acknowledge the hospitality and support from each other's respective institutes, during the course of this work.
}

\end{document}